\documentclass[aps,prl,superscriptaddress,twocolumn]{revtex4-1}

\usepackage{color}

\usepackage[utf8]{inputenc}
\usepackage[english]{babel}
\usepackage{amsmath,graphicx,enumerate}

\usepackage{amsmath,amsfonts}
\usepackage{graphicx}
\usepackage{siunitx}
\usepackage{booktabs,caption}
\usepackage{epsfig}
\usepackage{wrapfig}
\usepackage{subcaption}
\usepackage{hyperref}

\usepackage{amsthm}

\usepackage{cases}

\DeclareRobustCommand{\vect}[1]{
  \ifcat#1\relax
    \boldsymbol{#1}
  \else
    \mathbf{#1}
  \fi}

\begin{document}



\title{A handy fluctuation-dissipation relation to approach generic noisy systems and chaotic dynamics}


\author{M. Baldovin}
\affiliation{Dipartimento di Fisica, Universit\`a di Roma Sapienza, P.le Aldo Moro 5, 00185, Rome, Italy}
\author{L. Caprini}
\affiliation{Scuola di Scienze e Tecnologie, Universit\`a di Camerino - via Madonna delle Carceri, 62032, Camerino, Italy.\\ 
 }
\author{A Vulpiani} 
\affiliation{Dipartimento di Fisica, Universit\`a di Roma Sapienza, P.le Aldo Moro 5, 00185, Rome, Italy}

\date{\today}


\begin{abstract}
We introduce a general formulation of the fluctuation-dissipation relations (FDR) holding also in far-from-equilibrium stochastic dynamics. %
A great advantage of this version of the FDR is that it does not require the explicit knowledge of the stationary probability density function. Our formula applies to Markov stochastic systems with generic noise distributions: when the noise is additive and Gaussian, the relation reduces to those known in the literature; for multiplicative and non-Gaussian distributions (e.g. Cauchy noise) it provides exact results in agreement with numerical simulations. 
Our formula allows us to reproduce, in a suitable small-noise limit, the response functions of deterministic, strongly non-linear dynamical models, even in the presence of chaotic behavior: this could have important practical applications in several contexts, including geophysics and climate. As a case of study, we consider the Lorenz '63 model, which is paradigmatic for the chaotic properties of deterministic dynamical systems. 
\end{abstract}

\maketitle



\textit{Introduction --- } The Fluctuation-Dissipation relation (FDR) is surely one of the few pillars of non-equilibrium statistical mechanics: it allows to determine the statistical response of a dynamical system to small external perturbations in terms of  correlations of the unperturbed dynamics. The idea dates back to Einstein’s work on Brownian motion~\cite{E05} and further advance can be attributed to Onsager~\cite{O31}, who introduced the regression hypothesis for the relaxation of macroscopic perturbations.
The picture was subsequently unified by Kubo's theory which allows to predict the transport coefficients in terms of equilibrium correlations~\cite{K57}.
Since the FDR theory was initially derived for Hamiltonian systems slightly driven out of thermodynamic equilibrium by small perturbative forces, some confusion still persists on the range of applicability of these relations~\cite{sarracino2019fluctuation}. In fact, it is known that a \textit{generalized} FDR
\footnote{With the expression ``generalized FDR'' we refer to the identity relating the response function to time-correlations of suitable observables (evaluated in the absence of perturbation), generalizing the celebrated equilibrium FDR to generic systems including non-equilibrium dynamics.}
holds for a wider class of deterministic and stochastic systems going beyond the thermal equilibrium case~\cite{marconi2008fluctuation, maes}. 

The generalized   FDR is a topic of broad interest in statistical fluid mechanics and, starting from the contribution of Kraichnan~\cite{kraichnan1959}, it has been widely investigated in the context of turbulence, where it  plays a key role in the direct interaction approximation for the closure problem~\cite{K}. 
It also attracted the interest of the scientific community for its intriguing applications in geophysics and  climate, aiming to understand, e.g., atmospheric dynamics and oceanic currents~\cite{leith75climate, majda2010low, lucarini2017predicting, lembo2020beyond, ghil2020physics}.
In some of the above-mentioned applications, FDRs are formulated using approximate methods which allow to obtain the qualitative behavior of the response function due to the change of a relevant variable of the climate model under investigation. 
Among these techniques, the \textit{Classical quasi-Gaussian} approximation~\cite{leith75climate, abramov2007blended} is the simplest and most widely used one, but it fails when strong non-linearities are present. 
The \textit{Short-time FDR}~\cite{abramov2008new} can be instead applied to any system, even when the steady probability density function (p.d.f.) is unknown, but it is very unstable for chaotic deterministic dynamics. 
In the same framework, further advance was obtained with the so-called \textit{Hybrid Axiom A} method~\cite{ruelle1998general, abramov2007blended}, which provides more accurate results also for long times, again at the price of a Gaussian approximation.
The broad interest in the FDR emerged also in other areas of science 
ranging from equilibrium and non-equilibrium colloids~\cite{gomez2009experimental, maggi2010generalized}, to granular particles~\cite{puglisi2002fluctuation, bunin2008frequency,  gnoli2014nonequilibrium}, and even to biological systems, usually classified as active matter~\cite{fodor2016far, maggi2017memory, caprini2018linear}.
In these cases, generalized FDRs have been derived using near-equilibrium approximations~\cite{caprini2021generalized} and have been employed to provide expressions for the transport coefficients~\cite{dal2019linear}, the effective temperature~\cite{berthier2013non, szamel2017evaluating, petrelli2020effective} and the effect of mild shear~\cite{asheichyk2019response, Kruger2021Browian}. Exact relations for a colloid in an active environment have been recently investigated~\cite{maes2020fluctuating}.

First formulations of the generalized FDR were independently derived in the context of stochastic processes~\cite{A72} and chaotic dynamical systems~\cite{FIV90} under very general hypotheses. Since the knowledge of the steady-state probability distribution is required, in many practical cases they are rather difficult to use.
This hindrance can be overcome by alternative formulations of the generalized FDR derived through path-integral approaches, a method due to Furutsu~\cite{furutsu1964statistical} and Novikov~\cite{n65} dating back to the '60s and reformulated later on in the context of equilibrium field theory~\cite{cugliandolo1996large, andreanov2006dynamical, marguet2021supersymmetries}. 
The advantage of this technique, valid for systems driven by Gaussian noises, relies on the fact that it can provide an explicit formulation for the FDR (i.e. not depending on the p.d.f.) even in out-of-equilibrium conditions. Examples were studied in the context of spin systems~\cite{PhysRevE.78.041120} and non-equilibrium colloidal particles, both in overdamped~\cite{ss06, baiesi2009fluctuations, baiesi2009nonequilibrium, speck2009extended} and underdamped dynamics~\cite{baiesi2010nonequilibrium, sarracino2013time}. In these cases, FDRs have been expressed in terms of time-correlations involving the dynamical variables and their time-derivatives; recently, a version of the FDR only depending on time-correlations of dynamical variables have been derived for a specific active matter model~\cite{caprini2020fluctuation} and, more generally, for additive non-equilibrium stochastic processes~\cite{caprini2021generalized} also establishing an interesting connection with virial and equipartition theorems.

In this letter, using a different approach, not based on path-integral methods, we develop a unified formulation for the generalized FDR, holding in and out of equilibrium, even in the presence of multiplicative or non-Gaussian noises. 
Remarkably, our theory is able to reproduce, in a suitable small-noise limit, the response function also for strongly non-linear deterministic dynamical systems displaying chaotic behaviors.

\textit{A handy formulation of the FDR --- } 
Let us consider a rather general dynamical system described by state variables $\mathbf{x}=(x^{(1)}, x^{(2)}, ..., x^{(N)})$, which will be simply denoted as the \textit{unperturbed} system. 
If one introduces a small impulsive perturbation $\delta x^{(j)}(t)$ at time $t$ on the dynamics of the variable $x^{(j)}$, the trajectory of the system will slightly deviate to ${\bf x}'$ (\textit{perturbed} system). 
The response on the variable $x^{(i)}$ at time $t+\tau$ is  defined as:
\begin{equation}
\label{eq:response}
R_{ij}(\tau)=\frac{\overline{\delta x^{(i)}(t+\tau)}}{\delta x^{(j)}(t)}
\end{equation}
where the overline denotes the average over many repetitions of the experiment and $\delta x_i=x_i'-x_i$.
The formulation of the generalized FDR derived in~\cite{A72, FIV90} reads
\begin{equation}
\label{eq:fdr_vulp}
R_{ij}(\tau)= - \Big\langle x^{(i)}(t+\tau) { \partial \ln P_s({\bf x})  \over \partial x^{(j)}}\Big|_t\,   \Big\rangle\,,
\end{equation}
where the average $\langle \cdot \rangle$ is performed using the unperturbed system and $P_s({\bf x})$ is the stationary probability distribution of the unperturbed system (which is supposed to be non-vanishing almost everywhere).
Equation \eqref{eq:fdr_vulp} shows that the validity of the generalized FDR is not limited to equilibrium systems.
Beyond its  many conceptual advantages~\cite{marconi2008fluctuation}, however, Eq. \eqref{eq:fdr_vulp} has several practical limits, mostly emerging in systems with many degrees of freedom.
Indeed, the average on the right-hand side is pretty impossible to compute numerically, because $P_s({\bf x})$ is usually unknown except for a very few cases (e.g. gradient systems).

We focus now on discrete-time Markov processes of the form
\begin{equation}
\label{eq:discrete_time}
 \vect{x}_{t} =  \vect{x}_{t-\Delta t}+ \vect{f}(\vect{x}_{t-\Delta t})\Delta t+ \vect{\eta}_{t}\,,
\end{equation}
where $\vect{f}$ is a generic vector function with the physical meaning of a force, $\Delta t$ is a fixed time interval and 
the noise $\{\vect{\eta}_t\}$ is a sequence of random variables distributed according to 
some p.d.f. $\tilde{P}(\vect{\eta}_t|\vect{x}_{t-\Delta t})$. 
Let us notice that if the 
continuous-time limit $\Delta t \to 0$ exists, 
it can be written as a Langevin equation assuming It\^{o}'s convention. 
Since $\tilde{P}(\vect{\eta}_t|\vect{x}_{t-\Delta t})$ can be conditioned by the previous value of $\vect{x}$, in the continuous limit also multiplicative noises are allowed.
When the process~\eqref{eq:discrete_time} admits a stationary $P_s({\bf x})$, by invoking Markovianity one has
\begin{equation}
\label{eq:chapman}
\frac{\partial P_s(\vect{x}_t)}{\partial x^{(j)}_t}=  \int d \vect{x}_{t-\Delta t}\, P_s(\vect{x}_{t-\Delta t}) 
\frac{\partial}{\partial x^{(j)}_t} \mathcal{P}(\vect{x}_t | \vect{x}_{t-\Delta t})\,,
\end{equation}
where by $\mathcal{P}(\vect{x}_t | \vect{x}_{t-\Delta t})$ we denote the stationary propagator $\mathcal{P}(\vect{x}_t,t | \vect{x}_{t-\Delta t},t-\Delta t)$, dropping the explicit time-dependence for convenience of notation.
This propagator can be written in terms of the noise distribution as:
\begin{equation}
\label{eq:pdfcond}
 \mathcal{P}(\vect{x}_t | \vect{x}_{t-\Delta t})=\tilde{P}(\vect{x}_t-\vect{x}_{t-\Delta t}-\vect{f}(\vect{x}_{t-\Delta t})\Delta t \,|\,\vect{x}_{t-\Delta t})\,.
\end{equation} 
Substituting Eq.~\eqref{eq:chapman} and~\eqref{eq:pdfcond} into Eq.~\eqref{eq:fdr_vulp} (which also holds for discrete-time processes) one straightforwardly has
\begin{equation}
\label{eq:fdr_discr}
 R_{ij}(\tau)=-\left\langle x^{(i)}_{t+\tau} \frac{\partial}{\partial x^{(j)}_t}\ln \tilde{P}(\vect{\eta}_t(\vect{x}_{t},\vect{x}_{t-\Delta t})|\vect{x}_{t-\Delta t})\right\rangle\,,
\end{equation} 
where $\vect{\eta}_t(\vect{x}_{t},\vect{x}_{t-\Delta t})$ is obtained by inverting Eq.~\eqref{eq:discrete_time}.
More details on the derivation of Eq.~\eqref{eq:fdr_discr} are provided in the Supplemental Material (SM).

 The main advantage of the generalized FDR~\eqref{eq:fdr_discr} is that it only requires the knowledge of $\tilde{P}(\vect{\eta}_t|\vect{x}_{t-\Delta t})$, since $P_s$ does not explicitly appears as an observable in Eq.~\eqref{eq:fdr_discr} (its dependence is only implicitly contained in the average). In most cases $P_s$ is not known, and the above formulation provides a handy way to estimate response functions, even in the presence of stochastic dynamics ruled by multiplicative or non-Gaussian noises.
 This is also an advantage with respect to previous formulations based on path-integral approaches, such as those discussed in Refs.~\cite{ss06, seifert2010fluctuation}. Other, more general, formulations which include those cases are indeed present in the literature, but they are more involved and focused on the distinction between entropic and frenetic terms~\cite{baiesi2009fluctuations}, so that our results appear to be much simpler to use.
\begin{figure}
 \centering
 \includegraphics[width=\linewidth]{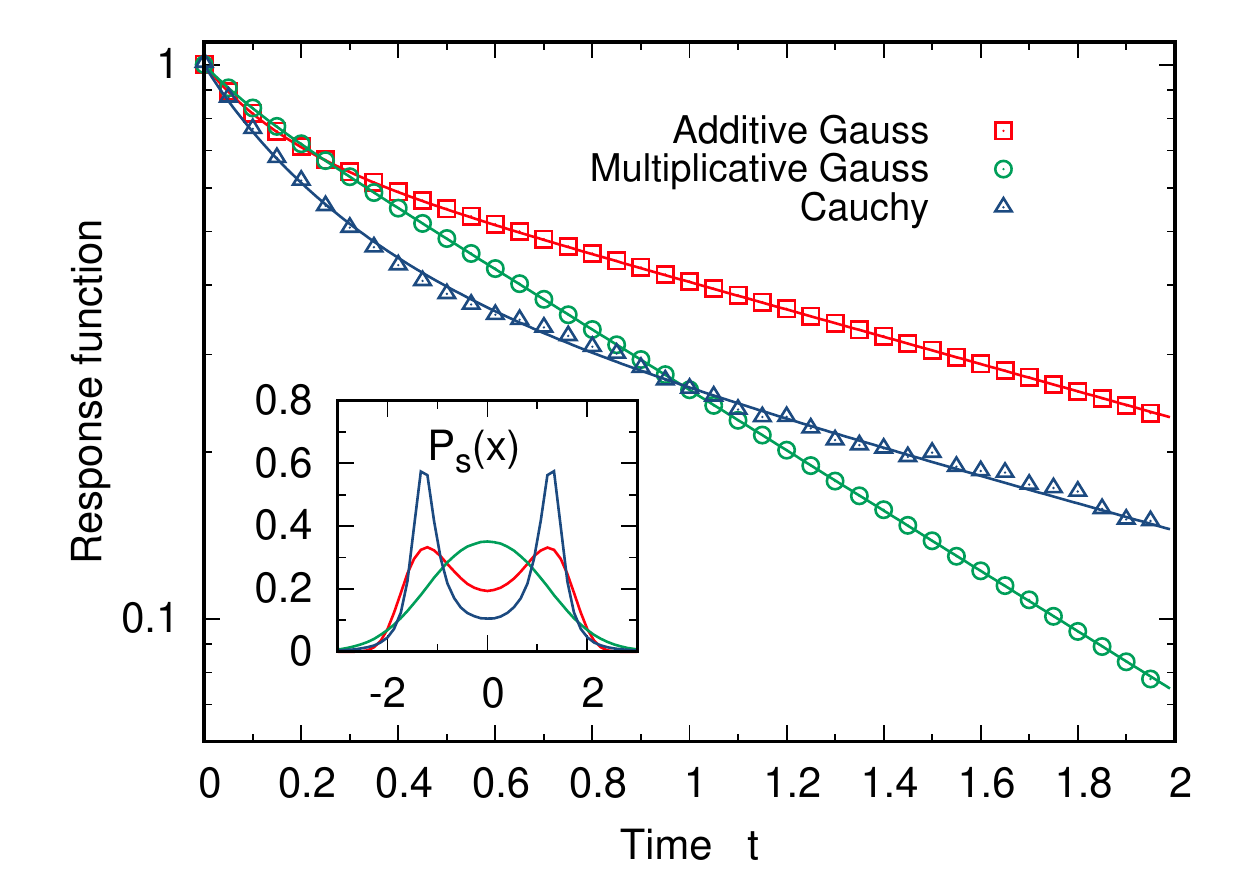}
 \caption{Response function, $R(\tau)$, and generalized FDR for system~\eqref{eq:discrete_time} in one dimension with $f(x)=-k(x^3-a x)$, induced by a double-well confining potential. Dynamics with additive Gaussian noise (red), multiplicative Gaussian noise (green) and Cauchy noise (blue), given by Eqs.~\eqref{eq:additive},~\eqref{eq:multiplicative} and~\eqref{eq:cauchy}, respectively, are considered. 
 Solid lines represent the $R(\tau)$ calculated by the definition~\eqref{eq:response} with $\delta x=10^{-2}$, while colored points are obtained using the FDR (Eq.~\eqref{eq:fdr_discr}).
 The other parameters used in the numerical study are: $k=1$, $a=3/2$,  $\sigma=1$, $D=1/2$ and $\Delta t=10^{-2}$.}
 \label{fig:noises}
\end{figure}
 
 To show the above point one can check the agreement between the actual response functions and the corresponding generalized FDRs for stochastic processes driven by noises with generic distributions.
 In particular, we consider Eq.~\eqref{eq:discrete_time} in the one-dimensional case, with a non-linear deterministic force $f(x)=-k(x^3-ax)$ (derived from a double-well potential) and noises described by three different statistics:
   \begin{subnumcases}{  \tilde{P}(\eta_t|x_{t-\Delta t}) \propto }
  \label{eq:additive}
\exp\left(-\frac{\eta_t^2}{4D\Delta t}\right)\\
\label{eq:multiplicative}
\exp\left(-\frac{\eta_t^2}{4D\Delta t(1+x_{t-\Delta t}^2)}\right)\\
\label{eq:cauchy}
   \frac{\Delta t}{\pi(\eta_t^2+\sigma^2\Delta t^2)}\,.
  \end{subnumcases}
Equations~\eqref{eq:additive} and~\eqref{eq:multiplicative} represent the cases of an additive and a multiplicative Gaussian noise. The form of the variance in Eq.~\eqref{eq:multiplicative} assures that for each value of $x_{t-\Delta t}$, $\tilde{P}(\eta)$ is well-defined.
Equation~\eqref{eq:cauchy} describes a Lorentzian distribution, corresponding to a non-Gaussian, Cauchy noise.
\footnote{Due to unavoidable computational limitations, in our simulations we had to impose a bound $|\eta_t|<M$ to the realization of the noise, i.e. we actually considered a \textit{truncated} Cauchy noise, in the spirit of Ref.~\cite{mantegna73stochastic}. For $M$ large enough we expect the corrections to be negligible, as suggested indeed by Fig.~\ref{fig:noises} where the analytical form of the FDR is computed according to the non-truncated p.d.f..}. 
This kind of noise is widely studied in the literature, in particular for its properties in continuous-time dynamics: indeed, it belongs to the class of so-called $\alpha$-stable noises which are typically used to model stochastic jump processes~\cite{ciesla2019multimodal, mantegna73stochastic}.
In Fig.~\ref{fig:noises} we show the response function $R(\tau)$ (the subscripts 
are suppressed for simplicity of notation), for the three cases, computed from 
Eq.\eqref{eq:response}, i.e. by directly perturbing the dynamics (see SM).
Let us note that, because of the presence of the external potential, one has a 
double time-decay regime for the response when additive Gaussian and Cauchy 
noises are considered, while a single regime is observed in the multiplicative 
Gaussian case. Consistently, the p.d.f. of $x$ is 
double-peaked in the former cases and it shows just one peak in the latter (see 
the inset of Fig.~\ref{fig:noises}). Although the qualitative behavior of the 
response crucially depends on the noise distribution, the comparison with the 
generalized FDR~\eqref{eq:fdr_discr} shows a remarkable agreement for all the
considered cases.


\begin{figure*}
 \centering
 \includegraphics[width=0.95\linewidth,keepaspectratio]{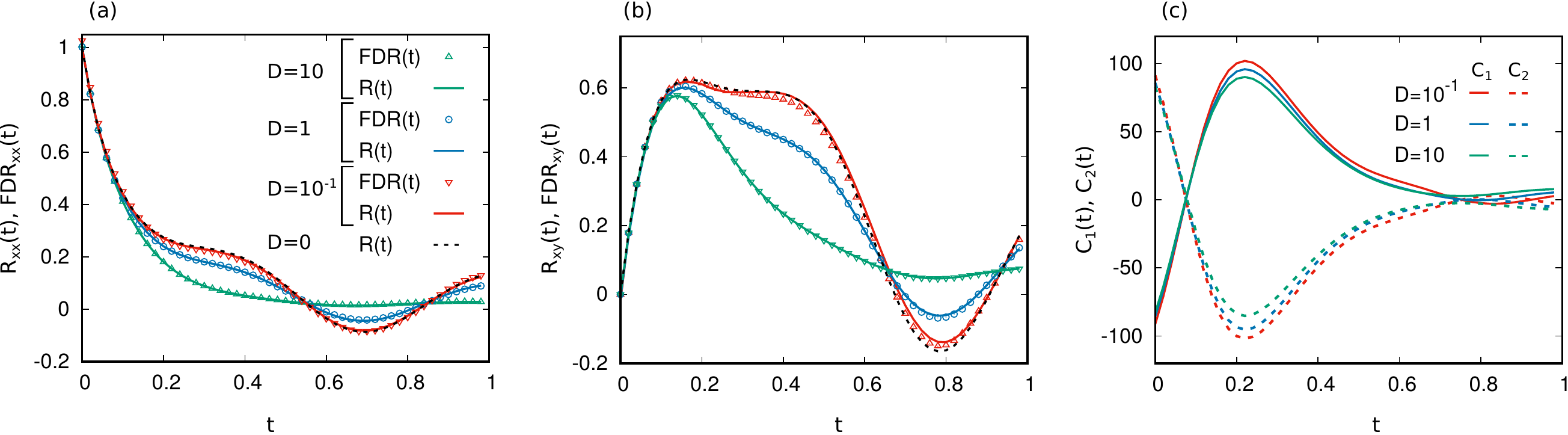}
 \caption{Response function and generalized FDR for the Lorenz '63 model. Panels~(a) and~(b): $R_{xx}(\tau)$ and $R_{xy}(\tau)$ (solid colored lines) obtained numerically by their definition, Eq.~\eqref{eq:response}, for different values of $D$.
 The dotted black line is obtained from the deterministic dynamics ($D=0$).
 Colored points are obtained from the FDR, namely Eq.~\eqref{eq:FDR_additivegaussian}. The legend in panel~(a) applies also to panel~(b).
 Panel (c) shows, for different values of $D$, the two time-correlations of the $xy$ element of the FDR, namely $C_1=\langle x(t)f_y(0) \rangle$ and $C_2=\langle f_x(t)y(0)\rangle$ (see Eq.~\eqref{eq:FDR_additivegaussian}).
 The parameters of the model are: $\sigma=10$, $\rho=28$ and $\beta=8/3$. The Langevin equation, i.e. the Lorenz model~\eqref{eq:lorenz63} with the noisy terms, is integrated using Heun's algorithm with time-step $\Delta t=10^{-3}$ and the impulsive perturbation $\delta x=10^{-1}$.
 }
 \label{fig:lorenz63}
\end{figure*}

For continuous-time processes Eq.~\eqref{eq:fdr_discr} becomes
\begin{equation}
\label{eq:fdr_cont}
 R_{ij}(\tau)=-\left\langle x^{(i)}(\tau) \frac{\partial}{\partial \dot{x}^{(j)}}\ln \tilde{P}(\dot{\vect{x}}-\vect{f}(\vect{x})|\vect{x})\Big|_0\right\rangle\,.
\end{equation} 
In particular, in the case of Gaussian additive noises the generalized FDR can be further simplified exploiting the stationarity of the process and it can be expressed as a time-correlation of the dynamical variables only (i.e., without involving their time-derivatives). In this way one recovers the relation already derived by one of us in~\cite{caprini2021generalized} (see also Ref.~\cite{matsumoto2014response} for an application in turbulent systems):
\begin{equation}
\label{eq:FDR_additivegaussian}
R_{ij}(\tau) = -\frac{1}{2 D_j} \left[ \langle x^{(i)}(\tau) f^{(j)}(0) \rangle + \langle f^{(i)}(\tau) x^{(j)}(0) \rangle \right] \,,
\end{equation}
where $D_j$ is the amplitude of the Gaussian noise $\eta_j$, here supposed diagonal for simplicity, satisfying $\langle \eta_i(t)\eta_j(s)\rangle = 2 D_j\delta_{ij}\delta(t-s)$.
Formula~\eqref{eq:FDR_additivegaussian} is consistent with earlier versions of equilibrium FDR  for the response function and it has been also applied in the context of non-equilibrium active matter models~\cite{caprini2021generalized}.

\textit{Application to chaotic systems: Lorenz '63 --- }
Eq.~\eqref{eq:FDR_additivegaussian} also finds applications in the context of non-linear dynamical systems, even with chaotic behaviors on strange attractors. 
We remark that our formula, being derived from  Eq.~\eqref{eq:fdr_vulp}, inherits the advantage of avoiding van Kampen's objection about the applicability to chaotic systems. 
Explicit expressions for the FDR in chaotic systems are particularly interesting for their applications in climate problems \cite{leith75climate,lucarini2017predicting}.
In this context, approximated approaches are usually employed to collect numerical results but they typically undergo strong limitations due to the chaotic nature of the dynamics~\cite{abramov2007blended,abramov2008new,cessac2007linear}.

As a test for Eq.~\eqref{eq:FDR_additivegaussian} in chaotic systems, we consider its application to the Lorenz '63 model, one of the most important and widely studied dynamics showing chaotic behavior~\cite{R1}. 
It was originally introduced as a simplified mathematical description for the atmospheric convection~\cite{lorenz1963deterministic}, playing a pivotal role to comprehend the typical mechanisms of climate problems. The evolution equations for this system read: 
\begin{subequations}
\label{eq:lorenz63}
\begin{align}
\dot{x}&=\sigma (y-x)\\
\dot{y}&=x (\rho-z) - y\\
\dot{z}&= xy -\beta z
\end{align}
\end{subequations}
where $\sigma=10$, $\rho=28$ and $\beta=8/3$ are fixed to display a chaotic behavior. The attractor is characterized by Lyapunov dimension $\approx 2.05$.
Studies on the response theory for this model were previously performed, in the spirit of the approximate methods for chaotic systems discussed in the introduction~\cite{Reick02Linear, lucarini2009evidence}.

It is known that in chaotic, deterministic, dissipative systems the stationary p.d.f. is singular and is typically characterized by a multifractal structure~\cite{R1}, so that Eq.~\eqref{eq:FDR_additivegaussian} (or Eq.~\eqref{eq:fdr_vulp}) cannot be applied in principle. 
We study therefore a stochastic version of Eqs.~\eqref{eq:lorenz63} in which we add a small source of additive Gaussian noise to the dynamics of each variable, namely $\sqrt{2 D} \boldsymbol{\eta}$, where $\boldsymbol{\eta}$ is a three-dimensional $\delta$-correlated white noise vector with zero average, and the coefficient $D$ rules the noise amplitude.  
Once these sources of noise have been included in the dynamics, the stationary p.d.f. is smoothed and we can employ the generalized FDR (also in its version Eq.~\eqref{eq:FDR_additivegaussian}). 
Figures~\ref{fig:lorenz63}(a) and~\ref{fig:lorenz63}(b) show two elements of the response function for the Lorenz dynamics~\eqref{eq:lorenz63}, namely $R_{xx}(\tau)$ and $R_{xy}(\tau)$, respectively. 
We compare the response computed according to the definition~\eqref{eq:response} with the generalized FDR~\eqref{eq:FDR_additivegaussian}: the good agreement between the two, for a broad range of values of $D$, confirms the exactness of the analytical relation, holding even for this highly non-linear dynamics with chaotic behavior.



\textit{The role of noise --- }
For small values of $D$ we expect the response functions to fairly approximate those obtained from the purely deterministic dynamics. Indeed, as already argued by Kolmogorov~\cite{R2}, the addition of a source of noise does not reflect on the chaotic nature of the dynamics, provided that the value of $D$ is small enough. 
In the absence of noise, a strange attractor shows a fractal structure at scales smaller than its typical size, and the Grassberger-Procaccia correlation behaves as $C(\ell) \sim \ell^{d_2}$ where $d_2$ is the correlation dimension~\cite{R1}. 
The presence of a noisy term in the evolution equation induces a smoothing of the attractor at scales $\ell < \ell_c$, where one has a trivial behavior $C(\ell) \sim \ell^d$  (with $d> d_2$ number of degrees of freedom of the system). One expects instead that the structure of the attractor at large scales (i.e. $\ell > \ell_c$) does not change if $D$ is small enough. The threshold $\ell_c$ depends of course on the strength of the noise: for instance, in the Lorenz model~\eqref{eq:lorenz63} we have verified that, in the presence of a noise with $D=10^{-1}$, the fractal structure of the attractor is unchanged for $\ell$ larger than $10^{-1}$.

In the light of the above discussion, it is natural to expect that the response function for deterministic dynamics should be arbitrarily close to the response (and, thus, to the generalized FDR) of the corresponding noisy case, with a suitably small choice of $D$. 
This is shown for both $R_{xx}(\tau)$ and $R_{xy}(\tau)$ in Fig.~\ref{fig:lorenz63}~(a) and~(b), where one observes that for decreasing values of $D$ the plotted curves approach their deterministic counterparts: see e.g. the comparison between red ($D=10^{-1}$) and black (deterministic case $D=0$) solid lines.
This means that Eq.~\eqref{eq:FDR_additivegaussian} (or Eq.~\eqref{eq:fdr_cont}) can also be employed to study the response function of deterministic systems, even in the presence of non-linearities and chaos.

Let us note that, since the response function are $O(1)$, the sum of the two time-correlations in the square brackets of Eq.~\eqref{eq:FDR_additivegaussian} must be $O(D)$.
In gradient systems an easy computation shows that both correlations are $O(D)$ (see for instance the results in Ref.~\cite{caprini2020fluctuation}). On the contrary, when the dynamics is chaotic and dissipative, an extended attractor is present and the fluctuations are mainly ruled by the deterministic dynamics. As a consequence, the two terms scale, in absolute value, with the size of the attractor, and their sum is $O(D)$ only by virtue of suitable cancellations.  
This is shown in Fig.~\ref{fig:lorenz63}~(c), where those contributions are plotted for different values of $D$ (only in the case of $R_{xy}(\tau)$ for simplicity). 
We remark that the smaller $D$, the larger the statistics one needs to obtain the same level of precision on the responses, due to the $1/2D_j$ prefactor in Eq.~\eqref{eq:FDR_additivegaussian}.

\textit{Final remarks --- }
Let us summarize the main message of this letter: for a generic stochastic system, even driven by a non-Gaussian noise, we introduce an exact generalized FDR which allows to write the response functions in terms of time-correlations in an explicit way, also in out-of-equilibrium conditions.
At variance with several other formulations of the generalized FDR, our approach only requires the evolution law of the dynamics and it is not based on the explicit knowledge of the steady-state probability distribution.
The method therefore allows us to investigate the behavior of dissipative chaotic systems: indeed, in the limit of vanishing noise, our generalized FDR is able to reproduce the response of deterministic systems. 

Among its many potential applications, our approach could be then adopted to characterize causal links among different variables in complex systems as climate dynamics and proteins. 
In particular, one may combine the ideas exposed in Ref.~\cite{baldovin2020causation} with the exact generalized FDR developed in this letter.

\vskip10pt
\begin{acknowledgments}
\textit{Acknowledgments --- } 
We thank A. Puglisi for useful discussions and a careful reading of the manuscript. This work is part of MIUR-PRIN2017 \textit{Coarse-grained description for non-equilibrium systems and transport phenomena (CO-NEST)}, whose financial support is acknowledged. 
All authors contributed equally to this work.
The authors declare no competing interests.
\end{acknowledgments}


\bibliographystyle{apsrev4-1}

\bibliography{bib}

\end{document}